\documentclass[journal=cgdefu, manuscript=article, layout=twocolumn]{achemso}

\usepackage{chemformula} 
\usepackage[T1]{fontenc} 
\SectionNumbersOn

\author{Si Wu}
\affiliation{Joint Key Laboratory of the Ministry of Education, Institute of Applied Physics and Materials Engineering, University of Macau, Avenida da Universidade, Taipa, Macao SAR 999078, China.}
\author{Yinghao Zhu}
\affiliation{Joint Key Laboratory of the Ministry of Education, Institute of Applied Physics and Materials Engineering, University of Macau, Avenida da Universidade, Taipa, Macao SAR 999078, China.}
\author{Haoshi Gao}
\affiliation{Joint Key Laboratory of the Ministry of Education, Institute of Applied Physics and Materials Engineering, University of Macau, Avenida da Universidade, Taipa, Macao SAR 999078, China.}
\alsoaffiliation{State Key Laboratory of Quality Research in Chinese Medicine, Institute of Chinese Medical Sciences (ICMS), University of Macau, Avenida da Universidade, Taipa, Macao SAR 999078, China.}
\author{Yinguo Xiao}
\affiliation{School of Advanced Materials, Peking University, Shenzhen Graduate School, Shenzhen 518055, China.}
\author{Junchao Xia}
\affiliation{Joint Key Laboratory of the Ministry of Education, Institute of Applied Physics and Materials Engineering, University of Macau, Avenida da Universidade, Taipa, Macao SAR 999078, China.}
\author{Pengfei Zhou}
\affiliation{Joint Key Laboratory of the Ministry of Education, Institute of Applied Physics and Materials Engineering, University of Macau, Avenida da Universidade, Taipa, Macao SAR 999078, China.}
\author{Defang Ouyang}
\affiliation{State Key Laboratory of Quality Research in Chinese Medicine, Institute of Chinese Medical Sciences (ICMS), University of Macau, Avenida da Universidade, Taipa, Macao SAR 999078, China.}
\author{Zhen Li}
\affiliation{Guangdong Provincial Engineering Research Center of Crystal and Laser Technology, Guangzhou, Guangdong 510632, China.}
\alsoaffiliation{Department of Optoelectronic Engineering, Jinan University, Guangzhou, Guangdong 510632, China.}
\author{Zhenqiang Chen}
\email{tzqchen@jnu.edu.cn}
\affiliation{Guangdong Provincial Engineering Research Center of Crystal and Laser Technology, Guangzhou, Guangdong, 510632, China.}
\alsoaffiliation{Department of Optoelectronic Engineering, Jinan University, Guangzhou, Guangdong, 510632, China.}
\author{Zikang Tang}
\email{zktang@um.edu.mo}
\affiliation{Joint Key Laboratory of the Ministry of Education, Institute of Applied Physics and Materials Engineering, University of Macau, Avenida da Universidade, Taipa, Macao SAR 999078, China.}
\author{Hai-Feng Li}
\email{haifengli@um.edu.mo}
\affiliation{Joint Key Laboratory of the Ministry of Education, Institute of Applied Physics and Materials Engineering, University of Macau, Avenida da Universidade, Taipa, Macao SAR 999078, China.}
\phone{+853 8822 4035}
\fax{+853 8822 2426}

\title{Super-Necking Crystal Growth and Structural and Magnetic Properties of SrTb$_2$O$_4$ Single Crystals}

\begin{document}

%
%


\begin{abstract}
We report on single-crystal growths of the SrTb$_2$O$_4$ compound by a super-necking technique with a laser-floating-zone furnace and study the stoichiometry, growth mode, and structural and magnetic properties by scanning electronic microscopy, neutron Laue, X-ray powder diffraction, and the physical property measurement system. We optimized the growth parameters, mainly the growth speed, atmosphere, and the addition of a Tb$_4$O$_7$ raw material. Neutron Laue diffraction displays the characteristic feature of a single crystal. Our study reveals an atomic ratio of Sr{:}Tb $ = 0.97(2){:}2.00(1)$ and a possible layer by layer crystal growth mode. Our X-ray powder diffraction study determines the crystal structure, lattice constants and atomic positions. The paramagnetic (PM) Curie{--}Weiss (CW) temperature $\theta_{\texttt{CW}} =$ 5.00(4) K, and the effective PM moment $M^{\texttt{eff}}_{\texttt{mea}} =$ 10.97(1) $\mu_\texttt{B}$ per Tb$^{3+}$ ion. The data of magnetization versus temperature can be divided into three regimes, showing a coexistence of antiferromagnetic and ferromagnetic interactions. This probably leads to the magnetic frustration in the SrTb$_2$O$_4$ compound. The magnetization at 2 K and 14 T originates from both the Tb1 and Tb2 sites and is strongly frustrated with an expected saturation field at $\sim$41.5 T, displaying an intricate phase diagram with three ranges.
\end{abstract}

\section{Introduction}

Magnetic frustration, as a result of competition between interactions that cannot be satisfied simultaneously, usually leads to fascinating magnetic properties such as spin liquid, spin ice, and cooperative paramagnetism in lanthanide-based compounds and continues to be a topic of considerable interest in condensed matter science as an excellent testing ground for theories and a potential for energy-related applications.\cite{Diep2004, Nakatsuji2006, Goremychkin2008, Castelnovo2008, Lee2008, Mila2015, Banerjee2017, Rau2019}

\begin{figure} [!t]
\centering \includegraphics[width=0.5\textwidth]{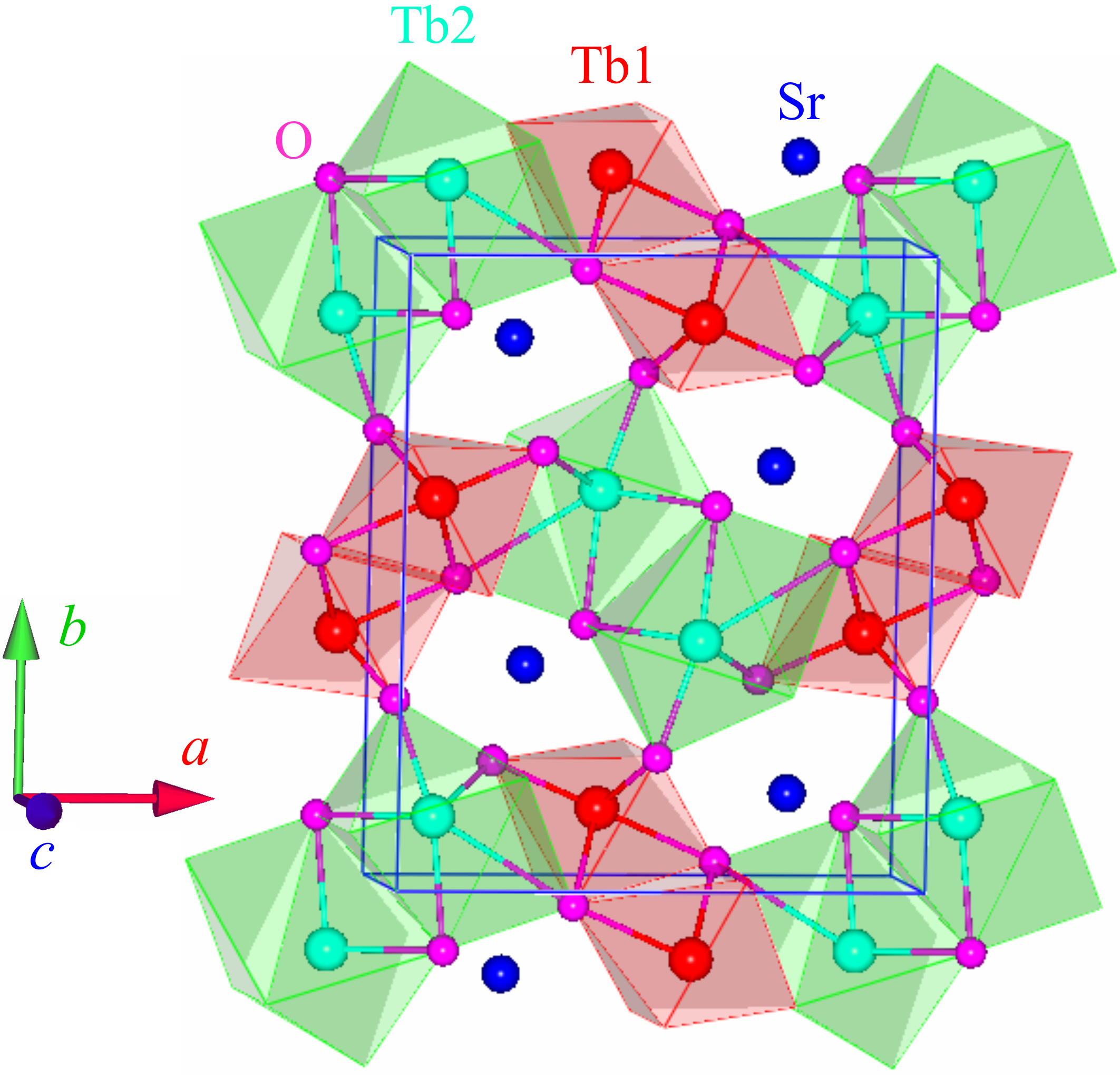}
\caption{Crystal structure (space group $Pnam$) of the single-crystal SrTb$_2$O$_4$ compound in one unit cell (solid lines) within the present experimental accuracy at room temperature. The Sr, Tb1, Tb2, and O ions are labeled. Within this structure, Tb1 and Tb2 ions form two kinds of TbO$_6$ octahedra. The four oxygen sites (O1, O2, O3, and O4) are all schematically illustrated with the same color code.}
\label{structure}
\end{figure}

In 1967, Barry and Roy for the first time synthesized the family of SrRE$_2$O$_4$ (RE = Y, Gd, Ho, Yb) compounds during exploring crystal chemical relationships between sesquioxides of certain rare earth ions and alkaline earth oxides with a subsidiary prediction based on simple field strength calculations.\cite{Barry1967} They established the existence of these new phases by employing standard X-ray diffraction techniques and determined the crystalline structure as a CaFe$_2$O$_4$-type orthorhombic structure.\cite{Barry1967} In 2005, Karunadasa et al. investigated the crystal structures, magnetic order, and susceptibility of polycrystalline SrRE$_2$O$_4$ (RE = Gd, Dy, Ho, Er, Tm, Yb) samples by neutron diffraction studies.\cite{Karunadasa2005} The authors reveal the existence of magnetic short-range orders down to $\sim$1.5 K and subsequently demonstrate that the compounds adopt the orthorhombic ($Pnam, Z =$ 4) structure (Figure~\ref{structure}). This structure accommodates two RE sites arranged in a honeycomb pattern with shared triangular edges and interconnections, which produces a strong geometric frustration for the magnetic ions.\cite{Karunadasa2005} This revitalizes the studies on related magnetically frustrated compounds in the following years.\cite{Balakrioshnan2009, Petrenko2014, Hayes2011, Fennell2014, Petrenko2008, Ghosh2011, Quintero2012, Hayes2012, Wen2015, Cheffings2013, Gauthier2017, Li2015Tm, Ouladdiaf2006, Paletta1968, Li2014-2, Aczel2014, Malkin2015, Aczel2015, Garlea2015, Bidaud2016, Gauthier2017-1, Gauthier2017-2, Fujimoto2017, Bobby2018, Jiang2018, Taikar2018, Mukherjee2018, Miao2019}

The single crystals of SrRE$_2$O$_4$ (RE = Y, Lu, Dy, Ho, Er) were grown by the floating zone technique using an optical mirror furnace\cite{Balakrioshnan2009, Li2014} for the subsequent neutron scattering studies.\cite{Petrenko2014} The observed neutron diffuse scattering in SrEr$_2$O$_4$ compound was ascribed to a ladder of Er triangles with a Monte Carlo simulation.\cite{Hayes2011} The crystal-field levels of SrHo$_2$O$_4$ and SrDy$_2$O$_4$ compounds were computed, which indicated a site-dependent anisotropic single-ion magnetism.\cite{Fennell2014} Neutron-scattering studies on SrRE$_2$O$_4$ (RE = Ho, Er, Yb) single crystals show their antiferromagnetic (AFM) transition temperatures of 0.62, 0.73, and 0.9 K, respectively, and generally demonstrate a coexistence of long- and short-range magnetic orders.\cite{Petrenko2008, Ghosh2011, Quintero2012, Hayes2012, Wen2015} The single crystalline SrDy$_2$O$_4$ compound shows a weak magnetic diffuse scattering only, and this short-range spin state persists down to $\sim$20 mK.\cite{Cheffings2013} One-third magnetization plateau was observed in the SrRE$_2$O$_4$ (RE = Dy, Ho, Er) compounds.\cite{Hayes2012, Wen2015, Gauthier2017} No magnetic ordering was observed down to $\sim$65 mK for the SrTm$_2$O$_4$ compound.\cite{Li2015Tm} Upon cooling, the SrYb$_2$O$_4$ single crystal undergoes a magnetic phase transition at $T_{\rm N}$ $\approx$ 0.9 K, forming a long-range commensurate noncollinear AFM order with a reduction in the ordered moment.\cite{Quintero2012}

The SrTb$_2$O$_4$ compound, as a member of the frustrated SrRE$_2$O$_4$ family, was first reported in 1968.\cite{Paletta1968} Previously, we performed a polarized and unpolarized neutron diffraction study,\cite{Li2014-2} determining that there exists an incommensurate AFM order with a transverse wave vector \textbf{Q}$_{\rm AFM}$ = (0.5924(1), 0.0059(1), 0) albeit with partially-ordered moments.\cite{Li2014-2} To our knowledge, there is no further study on the SrTb$_2$O$_4$ compound, probably due to the lack of large single crystals. Materials in the form of a single crystal are very useful for both scientific researches and industrial applications and can provide relatively reliable information of the structures and dynamics because they hold the translational symmetry over macroscopic distances.\cite{Kim2001, Triboulet2014, Naumov2015, Schmehr2017, Sun2018, Liu2018, Nandi2019, Balbashov2019, Konishi2019, Shen2019} To further shed light on the magnetic interaction and frustration necessitates larger and high quality SrTb$_2$O$_4$ single crystals.

In this paper, we optimized the growth parameters and grew cm-sized SrTb$_2$O$_4$ single crystals that are large enough for inelastic neutron scattering. During the crystal growth with our laser-diode-heated floating-zone (FZ) furnace, we managed the utilization of a super-necking technique, which is essential for obtaining the single crystals. We performed a preliminary in-house characterization on the structural and magnetic properties of the grown single crystals.

\begin{figure} [!t]
\centering \includegraphics[width=0.5\textwidth]{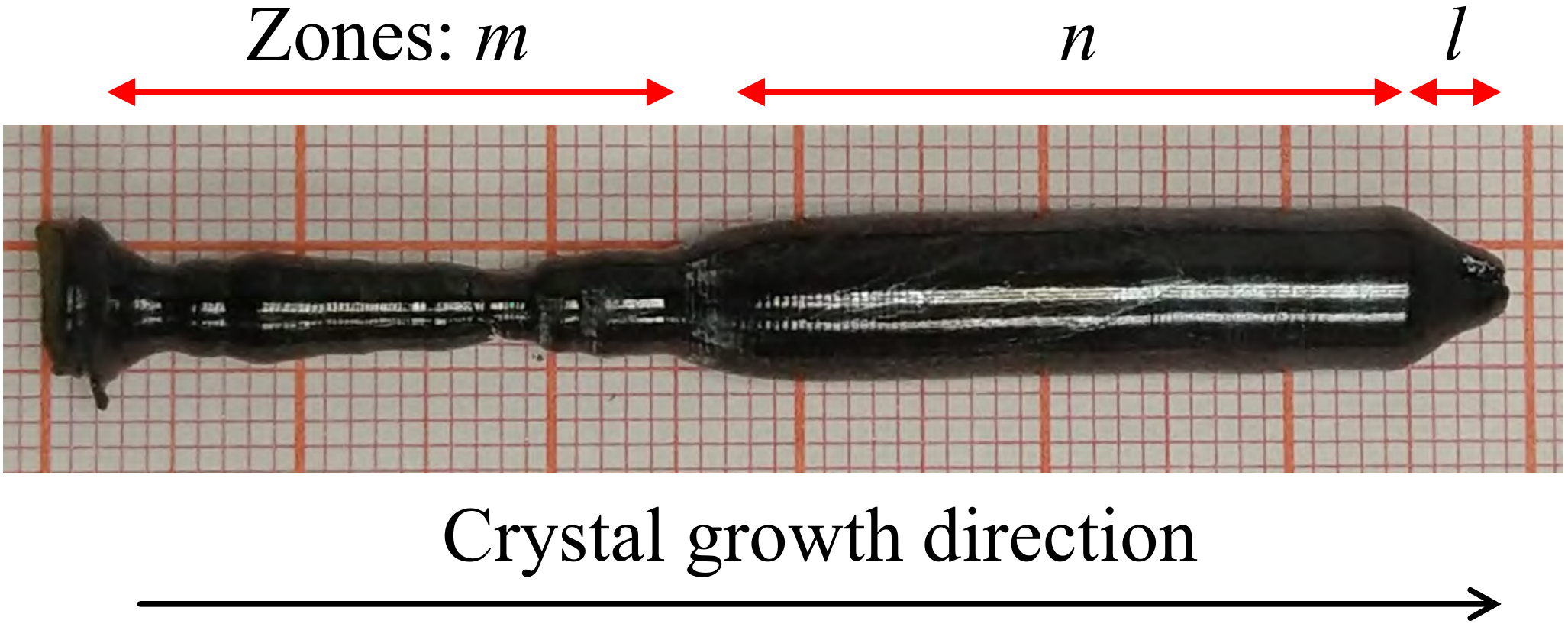}
\caption{One representative image of a SrTb$_2$O$_4$ crystal as grown by the Laser-diode floating-zone technique with a mixture of working gases of ($\sim$98\% Ar + $\sim$2\% O$_2$) under a pressure of $\sim$0.6 MPa. The zones \emph{m}, \emph{n}, and \emph{l} represent the super-necking regime, stable growth range, and final growth stage, respectively. The bottom arrow points out the crystal growth direction. During the final growth stage, we separated the seed rod from the molten zone step by step by manually adjusting the translation speeds of up and down shafts as well as reducing the heating power.}
\label{crystal}
\end{figure}

\begin{figure} [!t]
\centering \includegraphics[width=0.5\textwidth]{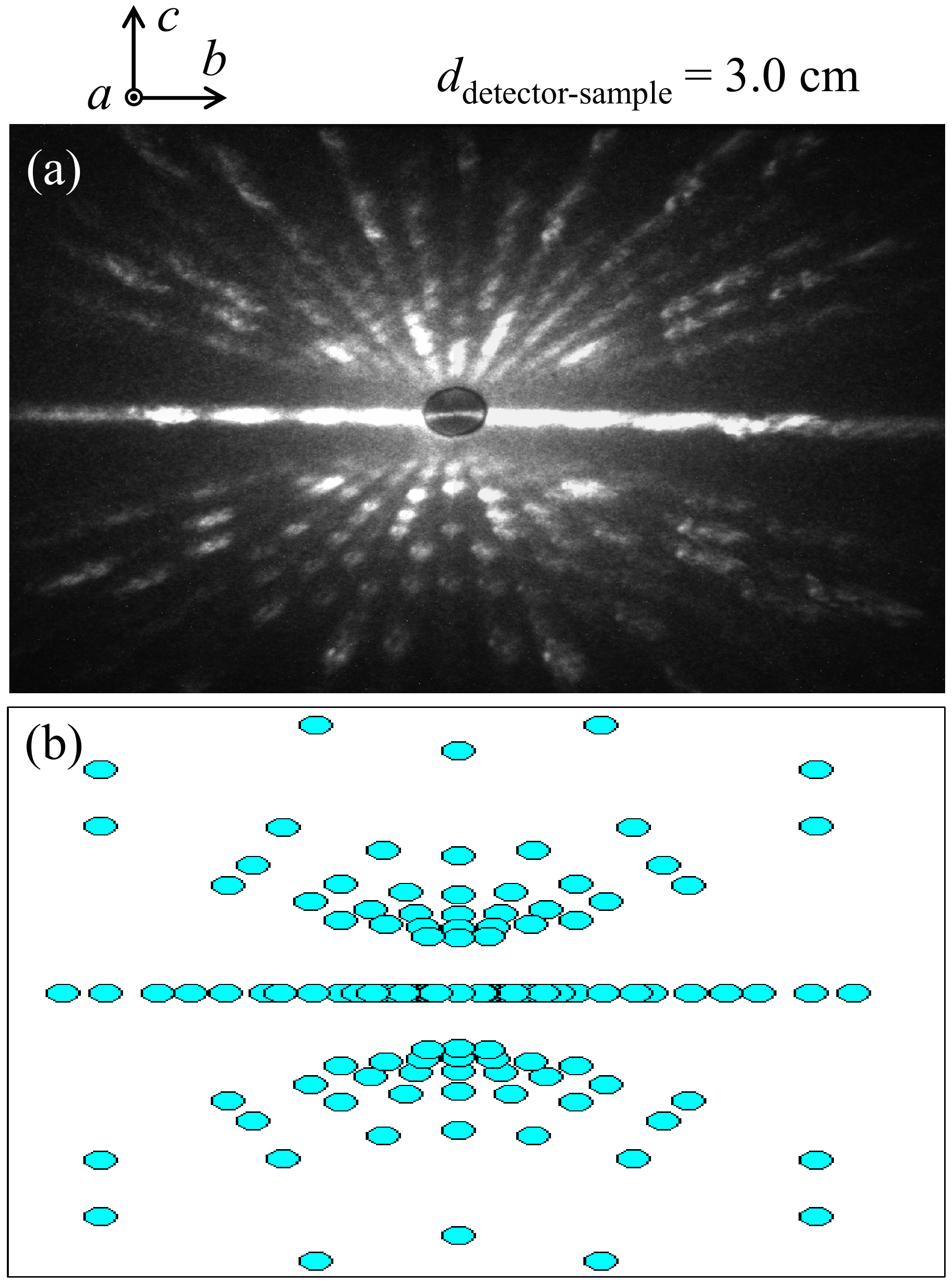}
\caption{
(a) One representative neutron Laue pattern of a SrTb$_2$O$_4$ single crystal with real-space lattice vectors as marked, collected with the $\texttt{"}$OrientExpress$\texttt{"}$ neutron Laue diffractometer,\cite{Ouladdiaf2006} located at ILL, France, with a distance of 3.0 cm between the detector and the sample. A pinhole with a 3 mm diameter was chosen for the neutron beam impinging on the crystal, and the exposure time was 3 min. The incoming neutron beam is perpendicular to the plane of the Laue spots and thus parallel to the unit-cell edge, \emph{i.e.,} the neutron beam is running along the crystallography \emph{a} axis. (b) Corresponding simulation of the neutron Laue pattern with the software $\texttt{"}$OrientExpress$\texttt{"}$.\cite{Ouladdiaf2006} Please refer to Ref. [23].}
\label{Laue}
\end{figure}

\section{Results and Discussion}

\subsection{Super-Necking Crystal Growth}

We first tried the crystal growth with a conventional FZ furnace (Crystal Systems Inc. Model FZ-T-10000-H-VI-VPO-PC). Although the furnace was equipped with four IR-heating halogen lamps as the heat source and four ellipsoidal mirrors as the reflectors, we found that its focused maximum temperature point was still below the melting temperature of the SrTb$_2$O$_4$ compound so that the rods could not be melted, and the crystal growth could not be performed. We thus turned to the laser diode FZ (LDFZ) furnace for the crystal growth. Compared with the conventional lamp-mirror-type FZ furnace, except for a much higher maximum temperature point, LDFZ furnace also holds a much steeper temperature gradient at the liquid{--}solid interface,\cite{Ito2013} which is more favorable for the nucleation and the formation of a stable growth state by easily controlling and attaining the balance between surface tension and hydrostatic pressure.\cite{Li2008} On the other hand, the sharp temperature gradient may result in cracks in grown crystals that are transparent or don't absorb the laser sufficiently. In our Quantum Matter Lab, the LDFZ furnace accommodates five 200 W laser diodes whose wavelength is 975(5) nm. The laser beam is in a line shape (width 4 mm $\times$ height 8 mm), and the working distance is $\sim$135 mm.

\begin{figure} [!t]
\centering \includegraphics[width=0.5\textwidth]{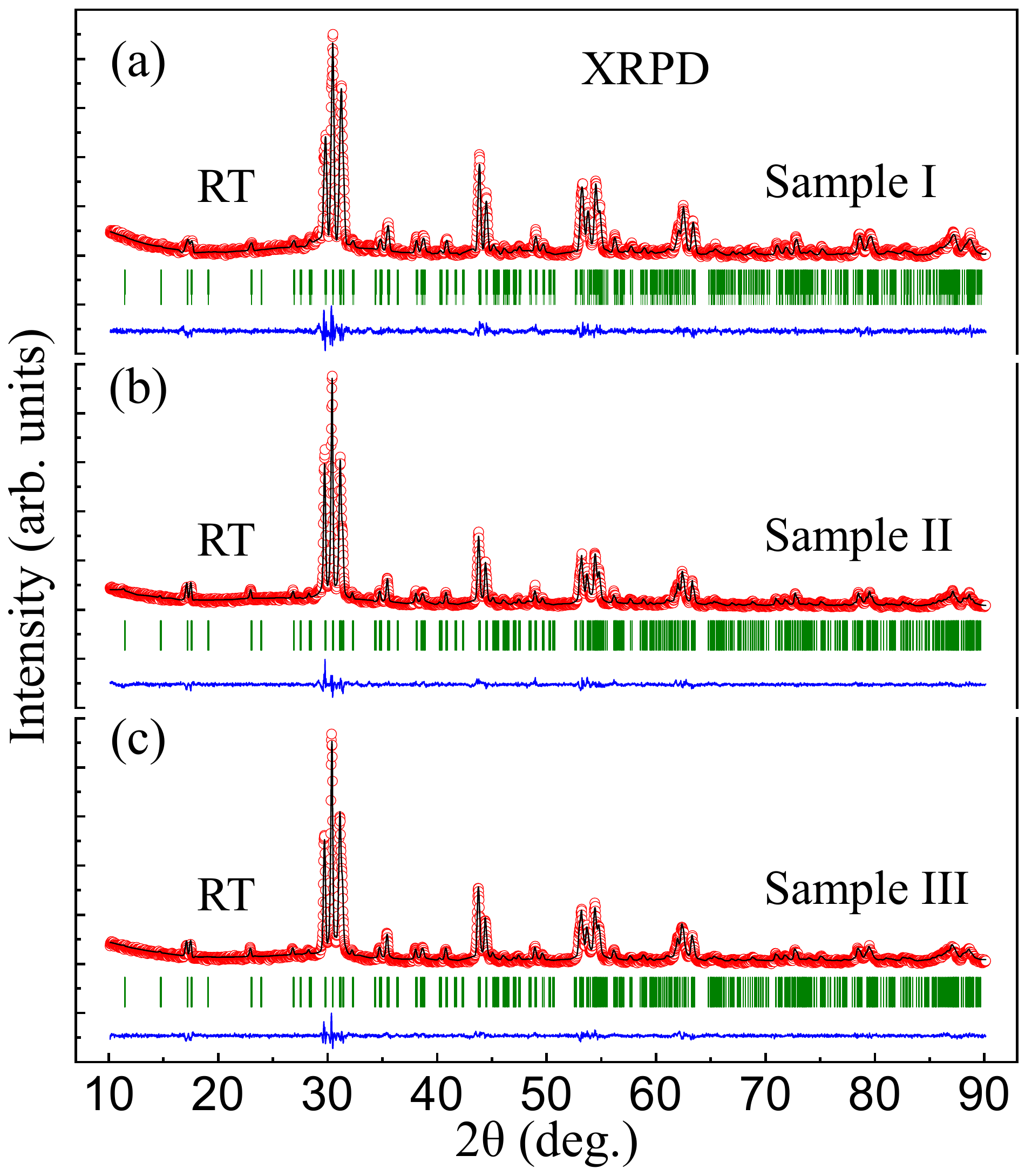}
\caption{Observed (circles) and calculated (solid lines) room-temperature (RT) in-house XRPD patterns of samples (a) I, (b) II, and (c) III, corresponding to the three parts \emph{l}, \emph{m}, and \emph{n} of the grown SrTb$_2$O$_4$ crystal (as marked in Figure~\ref{crystal}). The vertical bars mark the positions of nuclear Bragg reflections (space group $Pnam$). The lower curves represent the difference between observed and calculated patterns. The wavelengthes of copper $K_{\alpha1}$ (1.54056 {\AA}) and $K_{\alpha2}$ (1.544390 {\AA}) with a ratio of 2{:}1 were employed as the radiation.}
\label{XRD}
\end{figure}

During the early tentative growths, we observed some brown volatile matter (arising from the molten zone) depositing on the cold wall of the quartz tube. We tried to collect them; unfortunately, the amount was not enough for a detection with X-ray powder diffraction (XRPD). We speculated that they were from the evaporation of the Tb-related compounds. However, the possibility for the SrO/Sr(OH)$_2$ compounds cannot be ruled out. Therefore, the amount of the raw material Tb$_4$O$_7$ was optimized, and it was finally added 3--5\% more in compensation for the light evaporation during the process of single-crystal growth.

\begin{table*}[!t]
\caption{
Refined Structural Parameters of the Pulverized SrTb$_2$O$_4$ Samples I, II, and III from Zones \emph{l}, \emph{m}, and \emph{n} (Figure~\ref{crystal}), Respectively, Including Space Group $Pnam$, Lattice Constants, Unit-Cell Volume, Density (Calculated with the refinement), Atomic Positions, Isotropic Thermal Parameters (\emph{B}), and Goodness of Fit. The Wyckoff site of each ion was also listed. The numbers in parenthesis are the estimated standard deviations of the last significant digit.}
\label{structures}
\begin{tabular*}{0.88\textwidth}{@{\extracolsep{\fill}}lccccccccccccccccc}
\hline
\hline
                                      && I                && II              && III                                          \\
\hline
$a$ ({\AA})                           && 10.0910(3)       && 10.0955(5)      && 10.0983(3)                                   \\
$b$ ({\AA})                           && 11.9898(4)       && 11.9996(7)      && 11.9998(4)                                   \\
$c$ ({\AA})                           && 3.4463(1)        && 3.4489(2)       && 3.4506(1)                                    \\
$\alpha (\beta, \gamma)$
$(^\circ)$                            && 90               && 90              && 90                                           \\
$V$ ({\AA}$^3$)                       && 416.968(25)      && 417.814(42)     && 418.132(24)                                  \\
Density (g/cm$^3$)                    && 7.479(1)         && 7.464(1)        && 7.458(1)                                     \\
\hline
Sr(4$c$) \emph{x}                     && 0.7488(6)        && 0.7486(5)       && 0.7493(5)                                    \\
Sr(4$c$) \emph{y}                     && 0.6490(5)        && 0.6495(4)       && 0.6494(4)                                    \\
Sr(4$c$) \emph{z}                     && 0.25             && 0.25            && 0.25                                         \\
Sr(4$c$) \emph{B} (\AA $^2$)          && 2.47(1)          && 1.79(1)         && 2.13(9)                                      \\
\hline
Tb1(4$c$) \emph{x}                    && 0.4241(5)        && 0.4251(4)       && 0.4242(4)                                    \\
Tb1(4$c$) \emph{y}                    && 0.1111(3)        && 0.1120(3)       && 0.1118(3)                                    \\
Tb1(4$c$) \emph{z}                    && 0.25             && 0.25            && 0.25                                         \\
Tb1(4$c$) \emph{B} (\AA $^2$)         && 2.55(6)          && 2.64(6)         && 2.83(4)                                      \\
\hline
Tb2(4$c$) \emph{x}                    && 0.4173(5)        && 0.4184(4)       && 0.4184(4)                                    \\
Tb2(4$c$) \emph{y}                    && 0.6124(3)        && 0.6119(3)       && 0.6116(3)                                    \\
Tb2(4$c$) \emph{z}                    && 0.25             && 0.25            && 0.25                                         \\
Tb2(4$c$) \emph{B} (\AA $^2$)         && 2.55(6)          && 2.64(6)         && 2.83(4)                                      \\
\hline
O1(4$c$) \emph{x}                     && 0.2187(3)        && 0.2233(3)       && 0.2249(3)                                    \\
O1(4$c$) \emph{y}                     && 0.1891(2)        && 0.1694(2)       && 0.1856(2)                                    \\
O1(4$c$) \emph{z}                     && 0.25             && 0.25            && 0.25                                         \\
O1(4$c$) \emph{B} (\AA $^2$)          && 0.83(3)          && 2.64(3)         && 2.16(2)                                      \\
\hline
O2(4$c$) \emph{x}                     && 0.1318(2)        && 0.1467(2)       && 0.1493(2)                                    \\
O2(4$c$) \emph{y}                     && 0.4854(2)        && 0.4730(2)       && 0.4719(2)                                    \\
O2(4$c$) \emph{z}                     && 0.25             && 0.25            && 0.25                                         \\
O2(4$c$) \emph{B} (\AA $^2$)          && 0.83(3)          && 2.64(3)         && 2.16(2)                                      \\
\hline
O3(4$c$) \emph{x}                     && 0.4780(3)        && 0.4788(3)       && 0.4958(3)                                    \\
O3(4$c$) \emph{y}                     && 0.7905(2)        && 0.8076(2)       && 0.7949(2)                                    \\
O3(4$c$) \emph{z}                     && 0.25             && 0.25            && 0.25                                         \\
O3(4$c$) \emph{B} (\AA $^2$)          && 0.83(3)          && 2.64(3)         && 2.16(2)                                      \\
\hline
O4(4$c$) \emph{x}                     && 0.3935(2)        && 0.3867(2)       && 0.3941(2)                                    \\
O4(4$c$) \emph{y}                     && 0.4018(2)        && 0.3954(2)       && 0.4007(2)                                    \\
O4(4$c$) \emph{z}                     && 0.25             && 0.25            && 0.25                                         \\
O4(4$c$) \emph{B} (\AA $^2$)          && 0.83(3)          && 2.64(3)         && 2.16(2)                                      \\
\hline
$R_\texttt{p}$                        && 3.87             && 3.79            && 3.55                                         \\
$R_\texttt{wp}$                       && 4.96             && 4.84            && 4.49                                         \\
$R_\texttt{exp}$                      && 2.65             && 3.20            && 3.10                                         \\
$\chi^2$                              && 3.49             && 2.28            && 2.10                                         \\
\hline
\hline
\end{tabular*}
\end{table*}

\begin{figure} [!t]
\centering \includegraphics[width=0.5\textwidth]{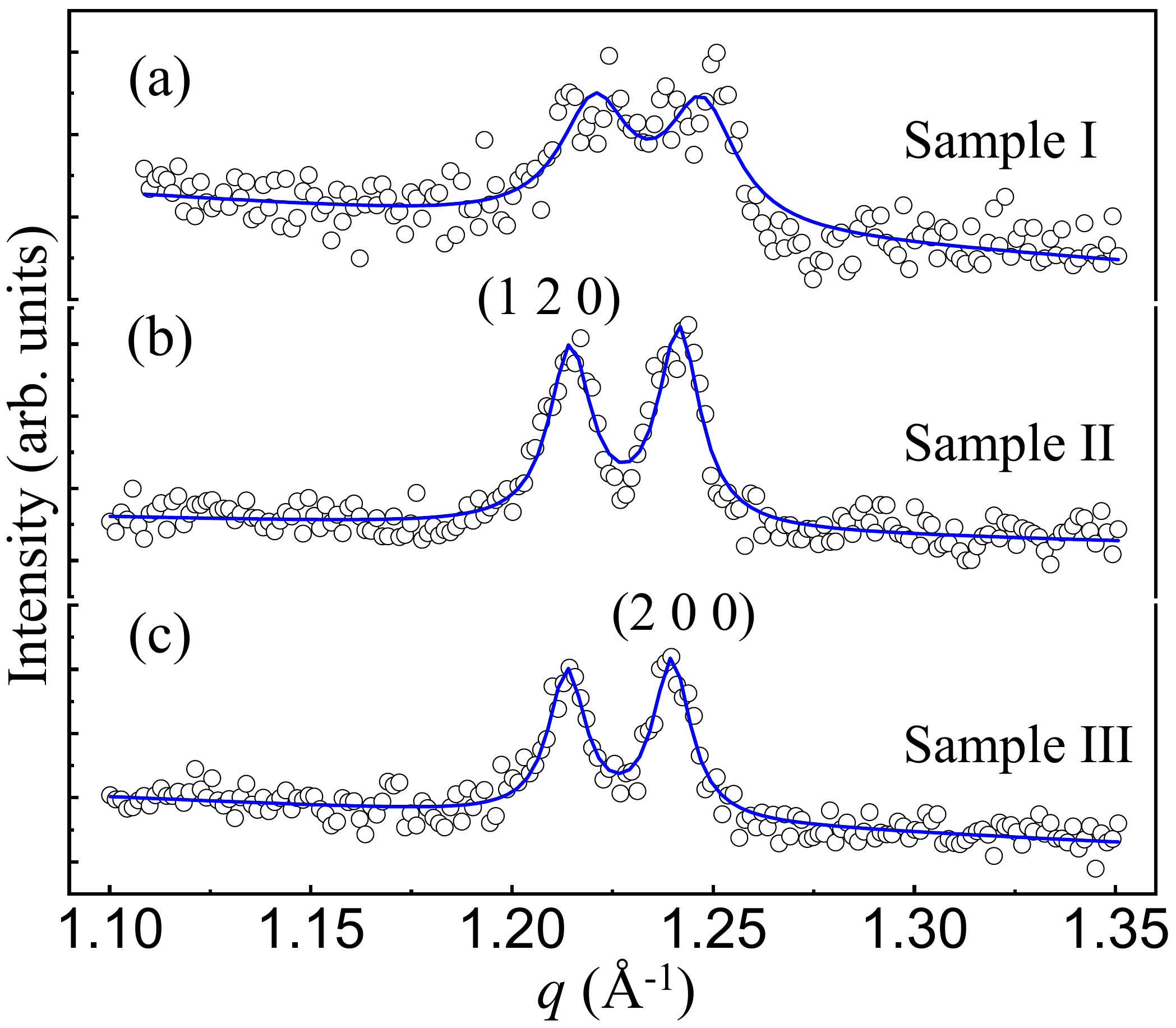}
\caption{XRPD data (circles) of the nuclear Bragg reflections of (1 2 0) and (2 0 0) in the \emph{q}-space of samples (a) I, (b) II, and (c) III. The data were measured at room temperature. The solid lines are two combined Lorentzian fits with equivalent full width at half maximum for the two peaks.
}
\label{peaks}
\end{figure}

\begin{figure} [!t]
\centering \includegraphics[width=0.5\textwidth]{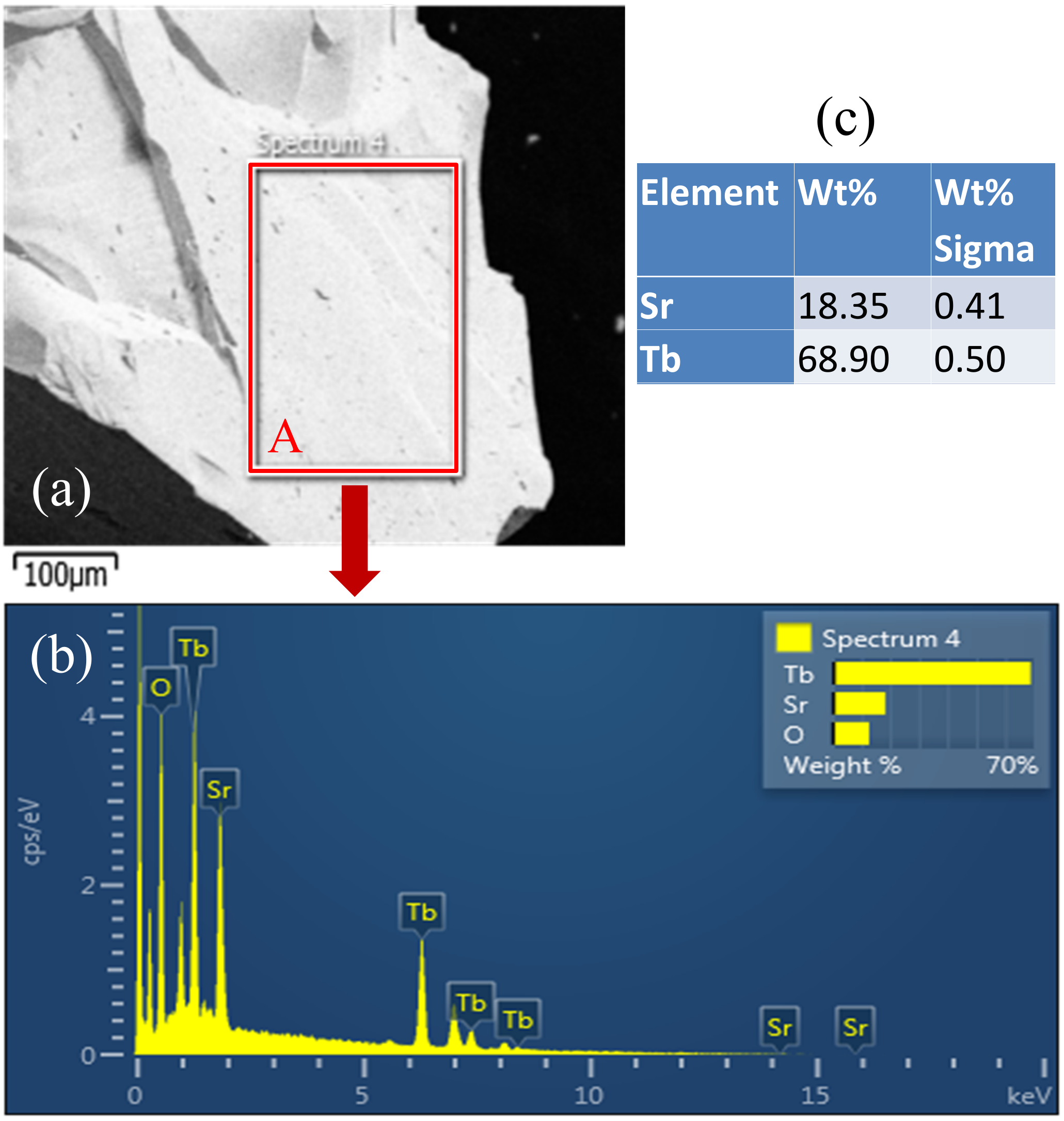}
\caption{(a) Scanning electronic microscopy image of a randomly selected small piece of broken SrTb$_2$O$_4$ crystal from the \emph{n} regime (as marked in Figure~\ref{crystal}). Scale bar represents 100 $\mu$m. (b) Energy-dispersive X-ray chemical composition analysis of the selected area A (as marked in panel (a)) of $\sim200 \times 300$ $\mu$m. (c) Extracted weight percentage (wt\%) of the chemical compositions of Sr and Tb elements as well as the corresponding error bars (wt\% sigma). It is pointed out that we ignore the composition analysis of oxygen because X-ray detection is not sensitive to the oxygen element at all.}
\label{SEMchenfen}
\end{figure}

In this study, the initial crystal growth was carried out on a polycrystalline seed rod. To diminish the number of spontaneous nucleation sites, taking the advantage of our LDFZ furnace, we used a super-necking technique for the crystal growth. As shown in Figure~\ref{crystal}, we divided the grown crystal into three regimes, zones \emph{m}, \emph{n}, and \emph{l}, as marked along the crystal growth direction. After connecting the feed rod with the seed rod and attaining a stable melting zone between them, we had held the state (left of zone \emph{m}) for approximately half an hour. Then, we tried to use the super-necking technique for growing the crystal with a lowering speed of 3--6 mm/h for the seed rod and 2--3 mm/h for the feed rod. During crystal growth within the zone \emph{m} ($\sim$2.3 cm), now and then, we manually adjusted the both translation speeds to keep the seed and feed rods conjoined with the melting zone and the grown part thinner than the feed rod because the crystallization speed exporting from the melting zone to the seed rod was larger than the solvent speed importing from the feed rod to the melting zone. The length of the \emph{m} zone is very long, $\sim$2.3 cm, and the corresponding diameter is $\sim$3--4 mm; therefore, we call this growth method as a super-necking technique. To our knowledge, it is extremely difficult to employ this technique with the conventional IR-heating FZ furnace. Between zones \emph{m} and \emph{n}, we maintained the translation speed of the seed rod at 6 mm/h and increased step by step the lowering speed of the feed rod up to 6 mm/h so that the diameter of the grown crystal became larger and larger (as displayed in the part between zones \emph{m} and \emph{n} in Figure~\ref{crystal}), gradually evolving into the stable growth range (zone \emph{n}). During the final growth stage (zone \emph{l}), while keeping the translation speed of the seed rod at 6 mm/h, we manually decreased the lowering speed of the feed rod as well as the heating power of the LDFZ furnace until the seed rod separated from the feed rod without collapsing the melting zone.

The setting of the initial growth conditions and the selection of the initial working pressure were from an empirical point of view. Finally, we optimized the growth parameters, mainly the growth speed, atmosphere, and the addition of the Tb$_4$O$_7$ raw material. Under the following conditions, we were able to smoothly grow the crystals: (i) The growth atmosphere was with flowing working gases of $\sim$98\% Ar plus $\sim$2\% O$_2$, controlled by the flowmeters equipped with the LDFZ furnace; (ii) The pressure of the working gases was approximately 0.6 MPa; (iii) The seed and feed rods rotated in opposite directions at 28 and 30 rotations per minute (rpm), respectively; (iv) The stable growth rate was fixed at 6 mm/h.

\subsection{Neutron Laue Study}

\begin{figure} [!t]
\centering \includegraphics[width=0.5\textwidth]{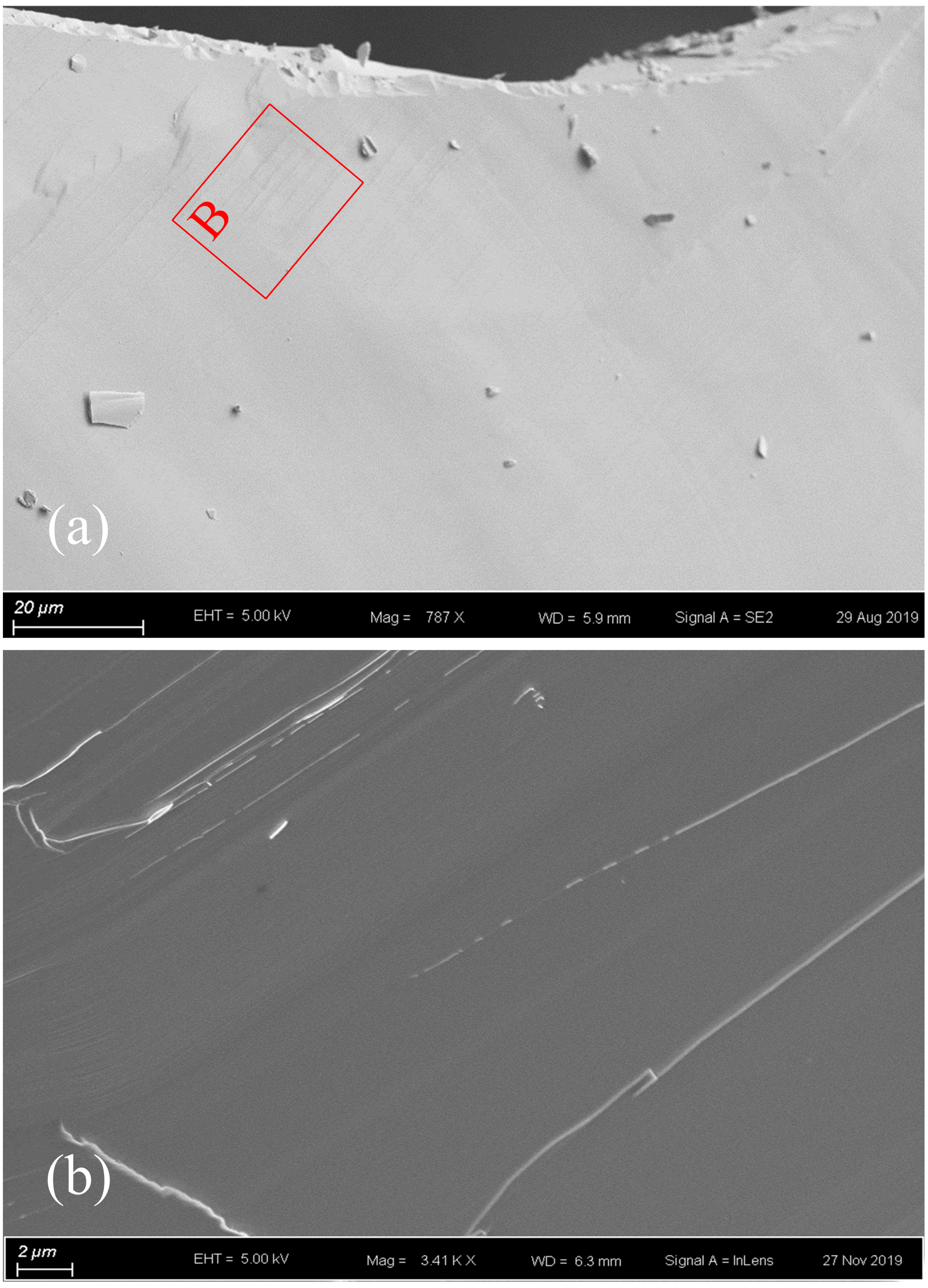}
\caption{Scanning electronic microscopy images of randomly selected pieces of the SrTb$_2$O$_4$ crystal from the \emph{n} regime (as marked in Figure~\ref{crystal}). (a) Scale bar represents 20 $\mu$m, and the total scanned area is $\sim22 \times 13.8$ $\mu$m. The magnification is 787$\times$. We marked one area named B (see detailed analysis in the text). (b) Scale bar represents 2 $\mu$m, and the magnification is 3410$\times$.}
\label{SEMzhigou}
\end{figure}

To confirm the single-crystalline nature of the grown crystals, we employed the neutron Laue diffractometer, OrientExpress (located at ILL, France), that was equipped with a CCD detector coupled to a neutron scintillator to monitor the Bragg diffraction spots in the reciprocal space of the SrTb$_2$O$_4$ compound. Figure~\ref{Laue}a shows the recorded back diffraction Laue pattern that maps the complete Miller indices of the [0, v, w] zone. The pattern is of two-fold symmetry with sets of Laue spots arranged in rings or lines about the center. We are confident about the \emph{c}-axis direction, vertical along the paper. However, within the present experimental accuracy, we cannot distinguish the \emph{a}-axis direction from that of the \emph{b}-axis because the lattice constants a and b are in close proximity to each other (as listed in Table~\ref{structures}).

We simulated the neutron Laue pattern with the software $\texttt{"}$OrientExpress$\texttt{"}$,\cite{Ouladdiaf2006} as shown in Figure~\ref{Laue}b, further endorsing our determination of the crystallographic orientations.

Our neutron Laue study together with the corresponding theoretical simulation verify the single-crystalline character of the grown crystals. It is pointed out that the crystallographic \emph{c} axis is $\sim$17$^\circ$ deviated from the crystal growth direction. This was determined by the goniometer of OrientExpress diffractometer as well as the software $\texttt{"}$OrientExpress$\texttt{"}$.\cite{Ouladdiaf2006} Similar neutron Laue patterns of a SrTm$_2$O$_4$ single crystal along three crystallographic axes were reported previously.\cite{Li2015Tm}

\subsection{Structural Characterizations}

To study the structural differences between grown crystals from the zones \emph{m}, \emph{n}, and \emph{l} (as marked in Figure~\ref{crystal}), we made three powdered samples I (from the zone \emph{l}), II (from the zone \emph{m}), and III (from the zone \emph{n}) for a controlled XRPD study at ambient conditions. We collected the XRPD patterns and refined them with the FULLPROF SUITE,\cite{Carvajal1993} as shown in Figure~\ref{XRD}. Within the present experimental accuracy, all collected Bragg reflections of the samples I, II, and III can be well indexed with the space group $Pnam$. The resulting crystal structure in one unit cell was displayed in Figure~\ref{structure} where the four oxygen sites (O1, O2, O3, and O4) were shown with the same color code. The refined structural parameters were listed in Table~\ref{structures}. It was noted that for samples I, II, and III, lattice constants (LCs) a, b, and c all display LC$^{\texttt{a}}_{\texttt{I}} <$ LC$^{\texttt{a}}_{\texttt{II}} <$ LC$^{\texttt{a}}_{\texttt{III}}$, LC$^{\texttt{b}}_{\texttt{I}} <$ LC$^{\texttt{b}}_{\texttt{II}} <$ LC$^{\texttt{b}}_{\texttt{III}}$, and LC$^{\texttt{c}}_{\texttt{I}} <$ LC$^{\texttt{c}}_{\texttt{II}} <$ LC$^{\texttt{c}}_{\texttt{III}}$. For example, for the LC a, 10.0910(3) {\AA} (sample I) $<$ 10.0955(5) {\AA} (sample II) $<$ 10.0983(3) {\AA} (sample III). These increases in LCs a, b, and c jointly lead to an enlargement of the unit-cell volume \emph{V}, 416.968(25) \AA$^3$ (sample I) $<$ 417.814(42) \AA$^3$ (sample II) $<$ 418.132(24) \AA$^3$ (sample III), as shown in Table~\ref{structures}.

To further explore the structural differences, we extracted the Bragg peaks of (1 2 0) and (2 0 0) of the three samples and exhibited them in Figure~\ref{peaks}. For the sample I, the measured XRPD data (circles) display a very broad peak without a clear splitting of the two peaks Figure~\ref{peaks}a. By comparison, we observed a distinct splitting of the two peaks for samples II Figure~\ref{peaks}b and III Figure~\ref{peaks}c. For a quantitative analysis, we fit simultaneously the measured data by two combined Lorentzian functions with equivalent full width at half maximum (FWHM) (solid lines), which resulted in FWHM$_\texttt{I} = $ 0.02075(3) \AA$^{-1}$ $>$ FWHM$_\texttt{II} = $ 0.01267(1) \AA$^{-1}$ $>$ FWHM$_\texttt{III} = $ 0.01114(1) \AA$^{-1}$. It is pointed out that the fits did not convolute the XRP diffractometer's resolution. We roughly calculated the charge-correlation length ($\xi$) by $\xi = \frac{2\pi}{{\texttt{FWHM}}}$, i.e., $\xi_\texttt{I} = 302.8(5)$ {\AA} $<$ $\xi_\texttt{II} = 496.1(5)$ {\AA} $<$ $\xi_\texttt{III} = 564.1(7)$ {\AA}. Therefore, sample III (from zone \emph{n} as marked in Figure~\ref{crystal}) possesses the best crystalline quality, which is due to that it has been grown from a stable growth state with all growth conditions fixed. During the final growth stage (zone \emph{l} as marked in Figure~\ref{crystal}), we reduced little by little the laser-heating power to smoothly separate the seed rod from the feed rod, which was probably the reason why the crystal (from zone \emph{l}) quality is the worst one.

It is extremely important that after reaching a stable crystal growth state, it would be better to fix all parameters for growing a crystal with a uniform crystallographic phase and a homogenous chemical stoichiometry.

\begin{figure} [!t]
\centering \includegraphics[width=0.5\textwidth]{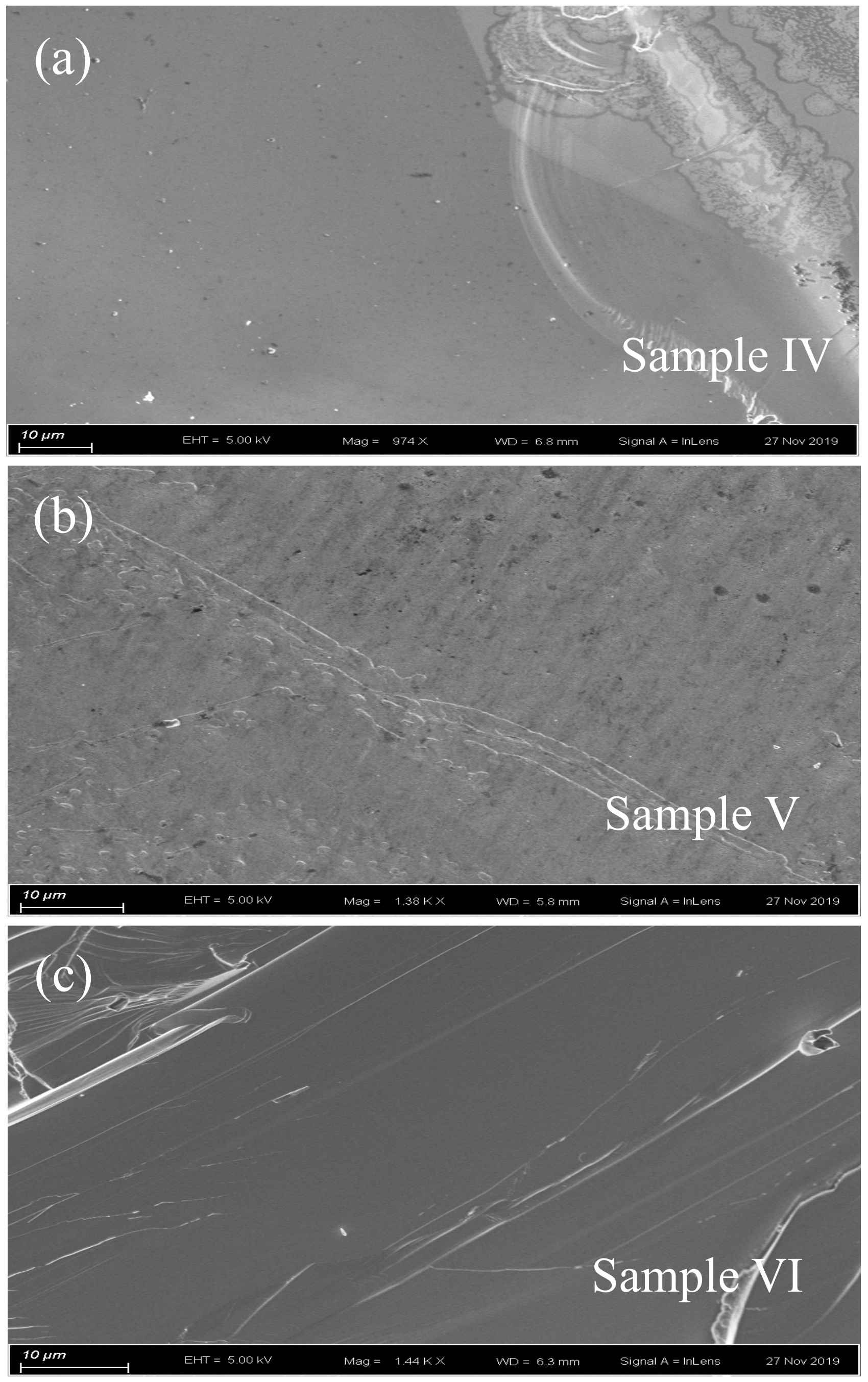}
\caption{Scanning electronic microscopy images of randomly selected pieces from (a) \emph{l} (sample IV) , (b) \emph{m} (sample V), and (c) \emph{n} (sample VI) regimes of the SrTb$_2$O$_4$ crystal (as marked in Figure~\ref{crystal}). Scale bar represents 10 $\mu$m.}
\label{lmn}
\end{figure}

\subsection{Scanning Electronic Microscopy}

We used the opinion of scanning electronic microscopy (SEM) to analyze the chemical compositions of one grown SrTb$_2$O$_4$ crystal. The measured sample was randomly selected from the broken SrTb$_2$O$_4$ crystal Figure~\ref{SEMchenfen}a. The detailed statistical analysis was performed on a flat surface with an area $\sim$200 $\times$ 300 $\mu$m (marked regime A in Figure~\ref{SEMchenfen}a). The energy-dispersive X-ray spectrum was displayed in Figure~\ref{SEMchenfen}b. The extracted weight percentage of the Sr and Tb elements were listed in Figure~\ref{SEMchenfen}c. It is pointed out that the X-ray technique is not sensitive to light oxygen ions. Based on this measurement, the extracted atomic ratio of Sr and Tb ions was Sr{:}Tb $=$ 0.97(2){:}2.00(1) when the Sr composition was normalized to the theoretical value (i.e., 2) of the element Tb. Supposing that the oxygen composition was stoichiometric, the chemical stoichiometry of the measured crystal was of Sr$_{0.97(2)}$Tb$_{2.00(1)}$O$_{4.00}$. This leads to 3.03(2) + for the oxidation state of the Tb element. The molar atomic ratio between Tb and Sr is close to the ideal one (i.e., 2). A more precise determination of the oxygen stoichiometry necessitates an inductively coupled plasma with optical emission spectroscopy analysis\cite{Li2015Tm} or a neutron diffraction study.\cite{Li2007-1, Li2007-2}

\begin{figure} [!t]
\centering \includegraphics[width=0.5\textwidth]{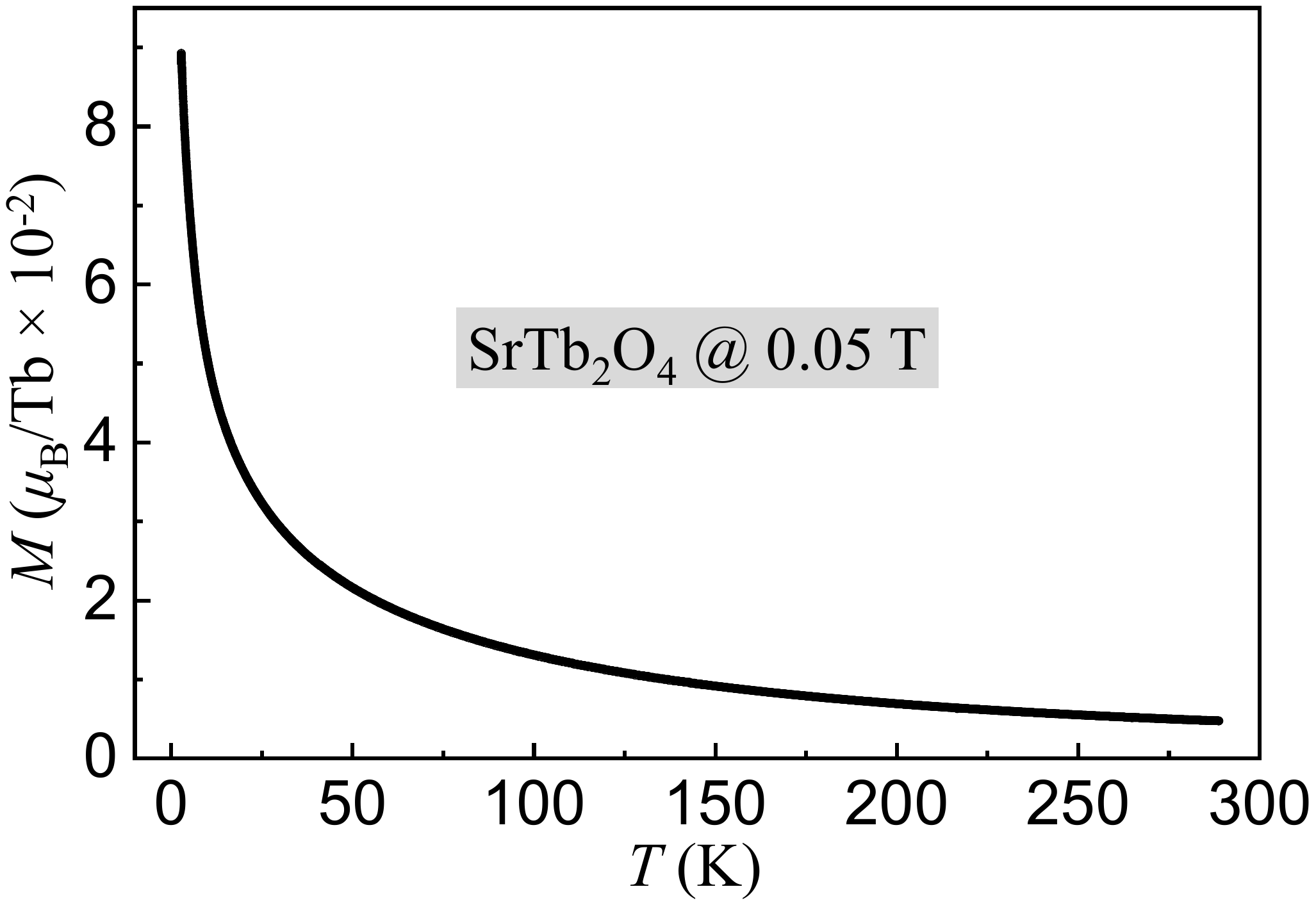}
\caption{Magnetization ($M$) per Tb ion in a randomly orientated SrTb$_2$O$_4$ crystal as a function of temperature, measured at an applied-magnetic field of 0.05 T. More than 9000 data points were recorded so that the error bars were embedded into the symbols.}
\label{MT}
\end{figure}

Figure~\ref{SEMzhigou}a shows the SEM image in an area $\sim$22 $\times$ 13.8 $\mu$m of a randomly selected piece from the \emph{n} regime of the SrTb$_2$O$_4$ crystal (as marked in Figure~\ref{crystal}). Albeit with some small fragments adhering to the surface, as a whole, the surface was very smooth, and we did not detect any clear grain boundaries, indicating a single crystal with good quality. With a close and careful check, we found some area (as the marked rectangle B in Figure~\ref{SEMzhigou}a) that was straightly wrinkled, as layered patterns of rocks. Such wrinkles are more obvious in a second piece of the SrTb$_2$O$_4$ crystal from the same \emph{n} regime, as shown in Figure~\ref{SEMzhigou}b. Based on this observation, we infer that the crystal growth mode of SrTb$_2$O$_4$ may be layer by layer, in agreement with the big difference between the lattice constant c and the lattice constants a and b (as listed in Table~\ref{structures}).

\begin{figure} [!t]
\centering \includegraphics[width=0.5\textwidth]{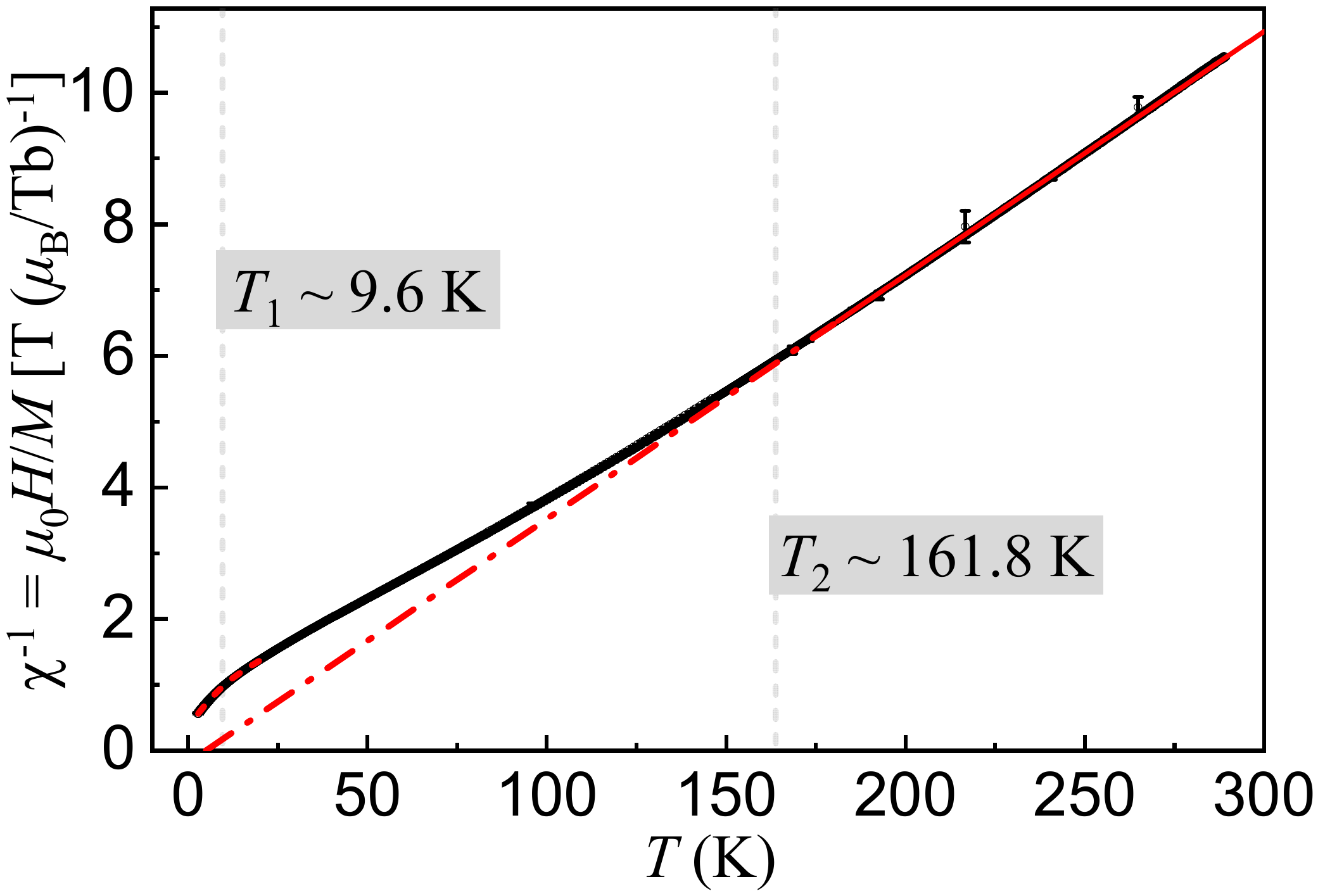}
\caption{Corresponding inverse magnetic susceptibility $\chi^{-1}$ (circles) of Tb ions in single-crystal SrTb$_2$O$_4$ versus temperature, deduced from Figure~\ref{MT}. The solid line shows a CW behavior of the data from 180 to 280 K, which was extrapolated to $\chi^{-1} = 0$ (dash-dotted line) to show the PM Curie temperature $\theta_{\texttt{CW}}$. The fit results are listed in Table~\ref{magP}. $T_2 \sim161.8$ K indicates the obvious deviation of the CW law behavior from measured data. We fit the measured data (from 2.8 to 20 K) with a combined lorentzian and linear function (dashed line) to obtain the anomaly of the measured data as marked at $T_1 \sim9.6$ K.}
\label{CWfit}
\end{figure}

Figure~\ref{lmn} shows SEM images of the grown crystals taken from the regimes of \emph{l} (a), \emph{m} (b), and \emph{n} (c) of the SrTb$_2$O$_4$ crystal (as marked in Figure~\ref{crystal}). Most part of sample IV (Figure~\ref{lmn}a) shows a single-crystalline pattern. Some polycrystalline forms still appear on the top-right corner of Figure~\ref{lmn}a. We ascribed this to the rapid cooling during the process of separating the seed rod from the feed rod. Sample V (Figure~\ref{lmn}b) seems to comprise a number of long grains, thus it may not be of single crystalline. This is consistent with the fact that sample V is from the super-necking growth regime, i.e., the initial stage of the crystal growth. Clearly, sample VI (Figure~\ref{lmn}c) from the \emph{n} regime of Figure~\ref{crystal} is a single crystal. Although samples IV, V, and VI display different forms of crystal, all are a single SrTb$_2$O$_4$ phase (Figure~\ref{XRD}). To figure out the difference between these three samples IV, V, and VI, as the foregoing analyses, we measured their energy dispersive X-ray spectra, and the extracted chemical stoichiometries are Sr$_{0.99(3)}$Tb$_{2.00(2)}$O$_{4.00}$ (sample IV), Sr$_{0.95(2)}$Tb$_{2.00(2)}$O$_{4.00}$ (sample V), and Sr$_{0.99(3)}$Tb$_{2.00(2)}$O$_{4.00}$ (sample VI). In Figure~\ref{SEMchenfen}, we determined the compositions of the sample from the \emph{n} regime of Figure~\ref{crystal} as Sr$_{0.97(2)}$Tb$_{2.00(1)}$O$_{4.00}$. There are no clear differences in the crystal-chemistry formulae, taking into account the error bars within the detecting resolution of the energy-dispersive X-ray spectra. Therefore, to understand the increase in the unit-cell volume as listed in Table~\ref{structures} necessitates elastic neutron scattering studies, with which the detailed bond lengths can be accurately determined taking into account the light oxygen element.

\subsection{Magnetic Properties}

\begin{figure} [!t]
\centering \includegraphics[width=0.5\textwidth]{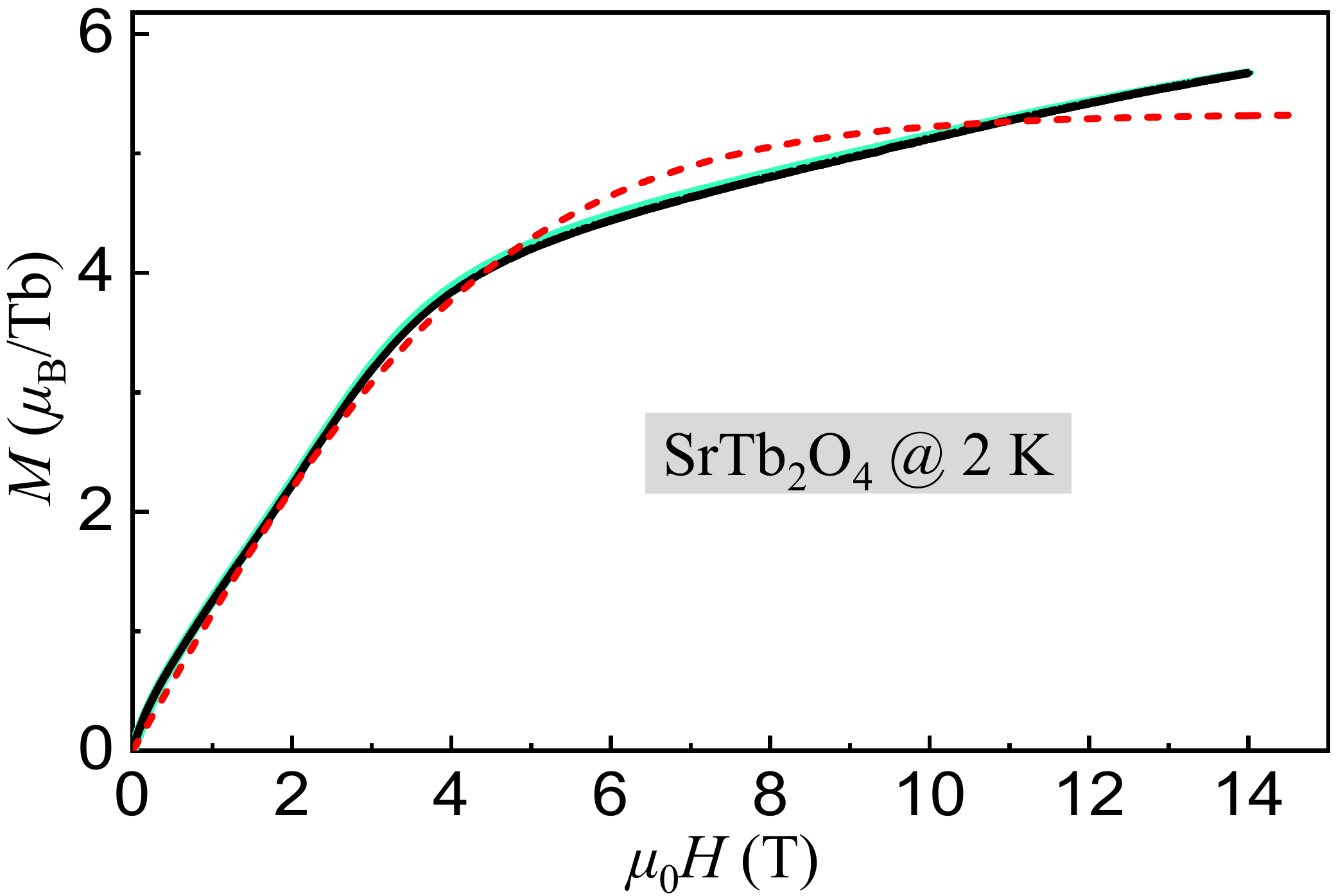}
\caption{\emph{M} measurements of a randomly orientated SrTb$_2$O$_4$ single crystal at 2 K with a loop of increasing (0 to 14 T) and decreasing (14 to 0 T) magnetic field (overlapped circles displaying like two lines). There is no clear difference between the two curves that nearly overlap together due to a high-density of data points ($\sim$5600). The short-dashed line is a fit to the Brillouin function (see detailed analysis in the text).}
\label{MH}
\end{figure}

\begin{figure} [!t]
\centering \includegraphics[width=0.5\textwidth]{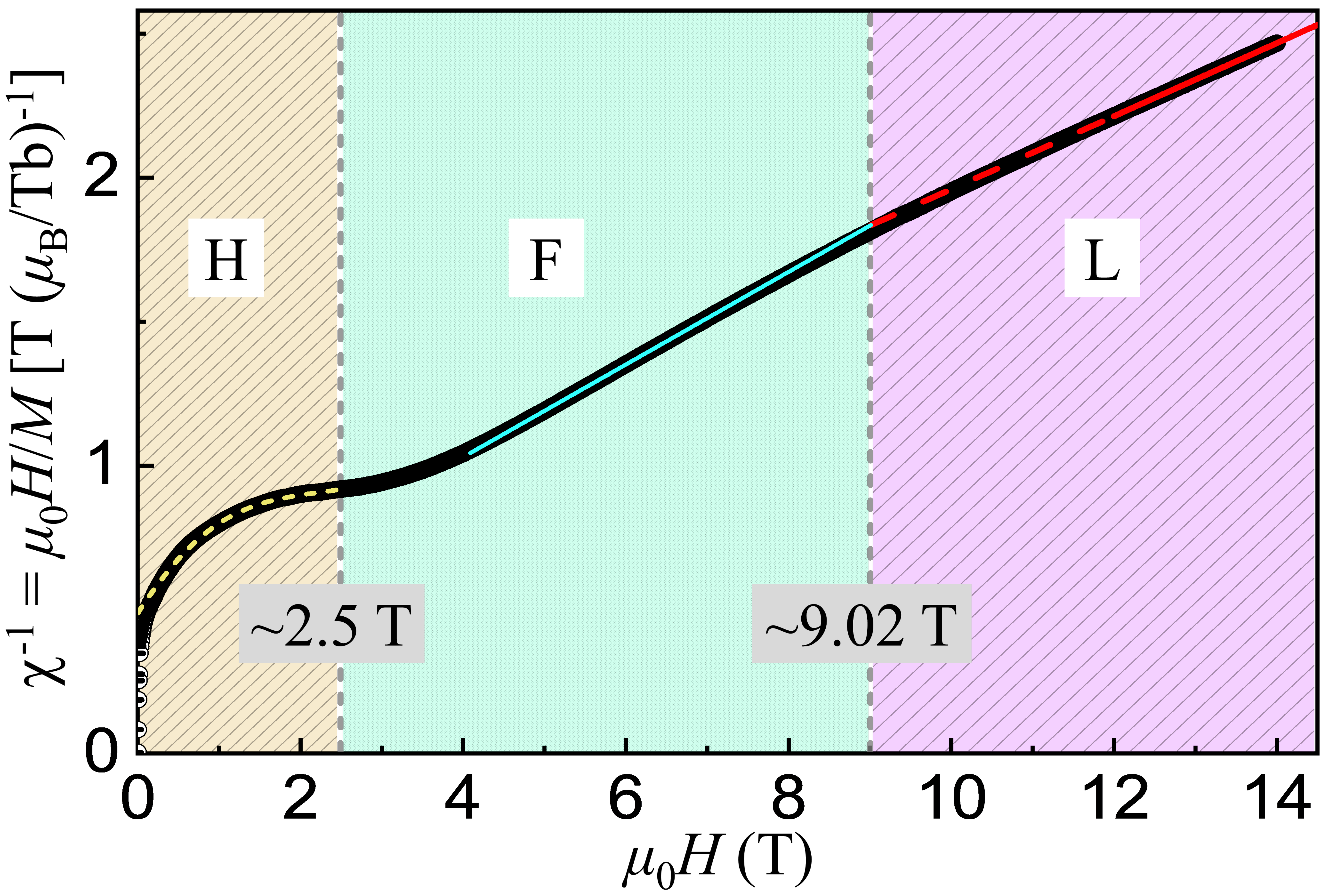}
\caption{Inverse magnetic susceptibility $\chi^{-1}$ ($\sim$5600 circles) of the single-crystal SrTb$_2$O$_4$ compound versus the applied-magnetic field $\mu_0H$ at 2 K. The solid, dashed, and short-dashed lines are fits to the measured data points in the respective field regimes. We divided the $\chi^{-1}$ versus $\mu_0H$ curve into three regimes: H (0--2.5 T), F (2.5--9.02 T), and L $(\geq 9.02$ T), to see detailed analyses in the text.}
\label{chi}
\end{figure}

Figure~\ref{MT} shows the magnetization measurement as a function of temperature at an applied-magnetic field $\mu_0H =$ 0.05 T. The magnetization was normalized to the Tb ions in the SrTb$_2$O$_4$ compound. Below $\sim$50 K it increases sharply, and above $\sim$50 K it keeps a smooth diminishment with an increase in temperature. As shown in Figure~\ref{CWfit}, we deduced the inverse magnetic susceptibility $\chi^{-1} = \frac{\mu_0H}{M}$ versus temperature. In the high-temperature pure paramagnetic (PM) state, the curve of $\chi^{-1}$ versus temperature can be well fit with a Curie{--}Weiss (CW) law:\cite{Li2014}
\begin{eqnarray}
\chi^{-1}(T) = \frac{3k_B(T - \theta_{\texttt{CW}})}{N_A M^2_{\texttt{eff}}},
\label{CWLaw}
\end{eqnarray}
where $k_B$ = 1.38062 $\times$ 10$^{-23}$ J/K is the Boltzmann constant, $N_A$ = 6.022 $\times$ 10$^{23}$ mol$^{-1}$ is the Avogadro's number, and $M_{\texttt{eff}}$ = $g_J \mu_\texttt{B} \sqrt{J(J + 1})$ is the effective PM moment.

The quantum numbers of the Tb$^{3+}$ ions in SrTb$_2$O$_4$ compound are listed in Table~\ref{magP}, where $J =$ 6 based on Hund{'}s rule for free Tb$^{3+}$ ions, $g_J = \frac{3}{2}$, the theoretical (theo) effective (eff) moment $M^{\texttt{eff}}_{\texttt{theo}} \approx$ 9.72 $\mu_\texttt{B}$, and the theo saturation (sat) moment $M^{\texttt{sat}}_{\texttt{theo}} = g_J J =$ 9.00 $\mu_\texttt{B}$.

As shown in Figure~\ref{CWfit}, we fit the deduced high-temperature data from 180 to 280 K with eq \ref{CWLaw} (solid line). By extrapolating the fit to $\chi^{-1} = 0$ (dash-dotted line), we obtained a PM CW temperature $\theta_{\texttt{CW}} =$ 5.00(4) K, implying a dominative ferromagnetic (FM) spin interaction. At $T =$ 161.8--288 K, the measured $\chi^{-1}$ data observe very well the CW law. Below $T_2 \sim$161.8 K, $\chi^{-1}$ data begin to go upward away from the CW law indicative of a formation of AFM spin correlations. Below $T_1 \sim$9.6 K, the measured $\chi^{-1}$ data try to turn back to the theoretical CW curve. That probably indicates an appearance of FM spin interactions. The coupling between AFM and FM interactions may result in an existence of magnetic frustration in the SrTb$_2$O$_4$ compound.

\begin{table}[!t]
\caption{
Quantum Numbers of the Single-Crystal SrTb$_2$O$_4$ Compound: Spin \emph{S}, Orbital \emph{L}, Total Angular Momentum \emph{J}, as well as the Land$\acute{\texttt{e}}$ Factor $g_J$ and the Ground-State Term $^{2S+1}L_J$. We also summarize the theoretical (theo) and measured (mea) values of the effective (eff), $M_{\texttt{eff}}$, and saturation (sat), $M_{\texttt{sat}}$, Tb$^{3+}$ moments, and the CW temperature, $\theta_{\texttt{CW}}$. The numbers in parenthesis are the estimated standard deviations of the last significant digit.}
\label{magP}
\begin{tabular*}{0.50\textwidth}{@{\extracolsep{\fill}}lr}
\hline
\multicolumn{2}{c} {SrTb$_2$O$_4$ single crystal}                                                                                                     \\
\hline
4$f$ ion                                                                                                 &              Tb$^{3+}$           \\
4$f^\texttt{n}$                                                                                          &              8                   \\
$S$                                                                                                      &              3                   \\
$L$                                                                                                      &              3                   \\
$J = L + S$ (Hund{\textquoteright}s rule for free Tb$^{3+}$)                                             &              6                   \\
$g_J$                                                                                                    &              3/2                 \\
$^{2S+1}L_J$                                                                                             &              $^7\texttt{F}_{6}$  \\
$M^{\texttt{eff}}_{\texttt{theo}} = g_{J} \sqrt{J(J+1)}$      $(\mu_\texttt{B})$                         &              $\sim$9.72          \\
$M^{\texttt{sat}}_{\texttt{theo}} = g_{J} J$ $(\mu_\texttt{B})$                                          &              9.00                \\
\hline
$M^{\texttt{eff}}_{\texttt{mea}}/\texttt{Tb}^{3+}$ (180--280 K, 0.05 T) $(\mu_\texttt{B})$               &              10.97(1)            \\
$\theta_{\texttt{CW}}$ (180--280 K, 0.05 T) (K)                                                          &              5.00(4)             \\
$M_{\texttt{mea}}/\texttt{Tb}^{3+}$ (2 K, 14 T) $(\mu_\texttt{B})$                                       &              5.68(1)             \\
\hline
\end{tabular*}
\end{table}

Based on our forgoing analysis, we also calculated the measured (mea) eff PM moment, i.e., $M^{\texttt{eff}}_{\texttt{mea}} =$ 10.97(1) $\mu_\texttt{B}$ per Tb$^{3+}$ ion. That $M^{\texttt{eff}}_{\texttt{mea}}$ value is larger than $M^{\texttt{eff}}_{\texttt{theo}} \approx$ 9.72 $\mu_\texttt{B}$, which may be ascribed to a formation of magnetic polarons.\cite{Li2012}

We also measured the magnetization versus applied-magnetic field up to 14 T at 2 K, as shown in Figure~\ref{MH}, where $M^{14\texttt{T}}_{2\texttt{K}} =$ 5.68(1) $\mu_\texttt{B}$ that is larger than half of $M^{\texttt{sat}}_{\texttt{theo}} =$ 9.00 $\mu_\texttt{B}$, implying a very strong magnetic frustration and that the magnetization is from both the Tb1 and Tb2 sites at a very high magnetic field. We schematically fit the curve of $M$ versus $\mu_0H$ with the Brillouin function.\cite{Zhu2019-2} Unfortunately, the fit (short-dashed line) cannot cover all features of the measured data. We fit linearly the magnetization at $\mu_0H =$ 13--14 T, i.e., $M = \chi \mu_0H + 3.98$, with magnetic susceptibility $\chi =$ 0.1211(2) $\texttt{T}^{-1} \mu_\texttt{B} \texttt{Tb}^{-1}$. Based on this fitting, we learn that reaching a complete saturated magnetic state requires an applied-magnetic field $\mu_0H \geq (M^{\texttt{sat}}_{\texttt{theo}} - 3.98)/\chi_{\textrm{13--14T}} \sim$41.5 T.

To deeply understand applied-magnetic field dependence of the magnetization, we extracted its inverse magnetic susceptibility $\chi^{-1}$, as shown in Figure~\ref{chi}. First, we fit the high-field (12--14 T) data with a linear function (slope = 0.12679(2) Tb$\mu_\texttt{B}^{-1}$) (solid line) and extrapolated the fit down to low field (dashed line). At $\mu_0H \approx$ 9.02 T, we observed the linear fit deviating upward from the measured data, which could be further demonstrated by a second linear fit (slope = 0.16087(1) Tb$\mu_\texttt{B}^{-1}$) to the data just below 9.02 T (solid line). The slope of the $\chi^{-1}$ curve is increased by $\sim$26.88\% for the two linear fits. We schematically fit the low-field data with a Gaussian function combined with a linear background contribution, shown as the short-dashed curve below 2.5 T. We can thus divide the measured data in Figure~\ref{chi} into three regimes H, F, and L based on the above analyses. In the regime H, as applied-magnetic field increases, the inverse magnetic susceptibility increases sharply at the beginning and then approaches step by step to a stable state at $\sim$2.5 T. During this process, the slope of the $\chi^{-1}$ curve decreases almost to zero. In the regime F, the slope of the $\chi^{-1}$ curve increases from zero to 0.16087(1) Tb$\mu_\texttt{B}^{-1}$ as $\mu_0H$ increases from 2.5 to 9.02 T, and then decreases to 0.12679(2) Tb$\mu_\texttt{B}^{-1}$ in the regime L (9.02--14 T). The results show a complicated magnetic field-dependent phase diagram of the SrTb$_2$O$_4$ compound.

\section{Conclusions}

To summarize, we have grown the SrTb$_2$O$_4$ single crystals by a super-necking technique with our laser-diode-heated FZ furnace. We utilized some tricks to make longer (up to 15 cm) and high-density (actual density $\geq$ 90\% of the theoretical one) seed and feed rods. The optimized crystal growth parameters are: $\sim$0.6 MPa working gases of ($\sim$98\% Ar + $\sim$2\% O$_2$), the seed (28 rpm) and feed (30 rpm) rods rotating oppositely, a growth speed of 6 mm/h, and 3--5\% additional inclusion of the raw Tb$_4$O$_7$ material. Neutron Laue measurements confirm that the grown crystals are of a single crystalline.

We studied comparatively three parts of the grown crystal by XRPD measurements. With FULLPROF refinements, we extracted quantitatively their detailed structural parameters (lattice constants, unit-cell volume, atomic positions, etc.) with the space group $Pnam$. We attributed the refined structural differences which indicate variations of composition in the crystal to the different growth parameters utilized during the process of crystal growth. Such growth parameters build stable conditions for the crystal growth. Present elemental analysis is not so accurate that we cannot correlate lattice parameters with composition. We determined the crystal compositions, Sr$_{0.97(2)}$Tb$_{2.00(1)}$O$_{4.00}$, by the SEM opinion of energy dispersive X-ray spectroscopy, providing that the oxygen composition is stoichiometric, and found a possible layer by layer growth mode.

By fitting the measured data of magnetization versus temperature with the CW law, we got a PM CW temperature $\theta_{\texttt{CW}} =$ 5.00(4) K and a measured effective PM moment $M^{\texttt{eff}}_{\texttt{mea}} =$ 10.97(1) $\mu_\texttt{B}$ per Tb$^{3+}$ ion. We divided the data into three ranges: (i) Between 161.8 and 288 K, they agree well with the CW law; (ii) At 9.6--161.8 K, AFM spin correlations form; (iii) Below $T_1 \approx$ 9.6 K, FM spin interactions probably appear. The coexistence of AFM and FM interactions leads to the magnetic frustration in the SrTb$_2$O$_4$ compound. The measured magnetization at 2 K and 14 T, $M^{14\texttt{T}}_{2\texttt{K}} =$ 5.68(1) $\mu_\texttt{B}$, is larger than half of $M^{\texttt{sat}}_{\texttt{theo}} =$ 9.00 $\mu_\texttt{B}$. This indicates that the magnetization originates from both the Tb1 and Tb2 sites at this magnetic field and the magnetism is strongly frustrated with an expected saturation magnetic field at $\sim$41.5 T. According to the different slopes, we classified the curve of the inverse magnetic susceptibility $\chi^{-1}$ versus the applied-magnetic field into three regimes: H (0--2.5 T); F (2.5--9.02 T); L (9.02--14 T).

Inelastic neutron scattering to reveal the spin-interaction parameters would be of great interest.\cite{Xiao2010, Zhang2013, Xiao2019} We paved the way for growing SrTb$_2$O$_4$ single crystals in this study. Comprehensive in-house characterizations and inelastic neutron scattering studies necessitate more SrTb$_2$O$_4$ single crystals with much higher quality. Presently, we are still working on how to improve the quality and size of SrTb$_2$O$_4$ single crystals.

\section{Experimental Techniques}

\textbf{4.1. Calcination of Polycrystalline Materials.} Polycrystalline samples of the SrTb$_2$O$_4$ compound were synthesized with raw materials SrCO$_3$ (Alfa Aesar, 99.99\%) and Tb$_4$O$_7$ (3--5\% more) (Alfa Aesar, 99.99\%) by a traditional solid-state reaction method.\cite{Li2008} The phase purity of raw materials was checked by XRPD at room temperature. To avoid the humidity effect on the quality of raw materials, they are always stored in a dry atmosphere at 493 K. They were quickly weighted at $\sim$473 K. The initial mixture of both raw materials was milled and mixed with a 50 mm-diameter ball by our Vibratory Micro Mill (FRITSCH PULVERISETTE 0) for 1 h, and then we calcined them in air for three times, 1373 K for 24 h , 1473 K for 36 h, and 1573 K for 24 h, with both heating-up and cooling-down ramps of 473 K/h. After each calcination, we milled and mixed the intermediate reaction materials again to get fine grains for a homogeneous reaction state.

\textbf{4.2. Longer and High-Density Seed and Feed Rods Preparation for Sintering and Crystal Growth.} To prepare longer and high-density feed rod is important and indispensable to a successful single crystal growth by the FZ method. It is usually hard to control the initial shape of the evacuated balloon. We utilized a range of techniques to overcome these difficulties. First, the rod shapes at its two ends are usually irregular, which makes the added force is insufficient, and the isostatic pressure cannot be uniformly transferred along the length of the rod. Before and after filling balloon with powder, we placed two flat Al-plates with a suitable diameter to force the shape of the two ends to be flat. With such a uniform diameter along the rod, the loading force along the rod direction can be increased largely and transferred easily. Second, we used a half Al-cylinder to fix the rod shape, before which it was stressed that we squeezed the enclosed powder so that its diameter was $\sim$2 mm larger than that of the Al-cylinder. Third, we used an isostatic pressurizer. With these simple but useful tricks, we were able to make longer (up to 15 cm) and high-density (actual density $\geq$ 90\% of the theoretical one) feed rods with any diameter required.

The product of the calcinations was filled into two plastic cylindrical balloons for preparations of seed and feed rods. After evacuating the air inside, we molded the balloons into cylindrical in shape, one diameter is $\sim$9 mm, and the other $\sim$6 mm. Both were hardened with a hydrostatic pressure of $\sim$70 MPa for about 20 min. The coating balloons were peeled off and removed. The prepared rods were sintered at 1573 K for 36 h in air. Finally, we obtained straight and uniformly-densified seed and feed rods with a homogenous composition distribution, which is essential for achieving and keeping a stable crystal growth state. We grew the single crystals of the SrTb$_2$O$_4$ compound with the FZ technique\cite{Li2008, Li2009-1} using a laser diode FZ furnace (Model: LD-FZ-5-200W-VPO-PC-UM)\cite{Zhu2019-1} well-equipped at the University of Macau, Macao, China.

\textbf{4.3. X-Ray Powder Diffraction Measurement.} We gently ground the single crystals of the SrTb$_2$O$_4$ compound into powder samples and characterized them by XRPD from 2$\theta =$ 10 to 90$^\circ$ with a step size of 0.02$^\circ$ on an in-house diffractometer (Rigaku, SmartLab 9 kW), employing both the copper $K_{\alpha1}$ (1.54056 {\AA}) and $K_{\alpha2}$ (1.54439 {\AA}) with a ratio of 2{:}1 as the radiation, in a Bragg-Brentano geometry at a voltage of 45 kV, a current of 200 mA, and ambient conditions.

\textbf{4.4. Structural Refinement.} We refined the collected XRPD data with the software FULLPROF SUITE.\cite{Carvajal1993} We chose the Pseudo-Voigt function for modeling the Bragg peak shape. We used a linear interpolation between automatically-selected data points to estimate the background contribution. We refined the parameters of the scale factor, zero shift, peak shape parameters, asymmetry, lattice parameters, atomic positions, and isotropic thermal parameters. In the final step of analysis, all parameters were refined together.

\textbf{4.5. Scanning Electron Microscope Measurement.} The SEM images and the energy dispersive X-ray chemical composition analysis of randomly selected SrTb$_2$O$_4$ crystals were carried out with the ZEISS Sigma. All crystals were sputter-coated with gold to make them electrically conductive. SEM studies were performed under an accelerating voltage of 5.00 kV in vacuum.

\textbf{4.6. Neutron Laue Measurement.} We measured neutron Laue pattern with a $\sim$3 g cleaved piece of the SrTb$_2$O$_4$ single crystal using the neutron Laue backscattering diffractometer, OrientExpress, located at the Institut Laue-Langevin (ILL), Grenoble, France.

\textbf{4.7. Magnetization Measurement.} We carried out the dc magnetization measurements, from 2.8 to 288 K at an applied-magnetic field $\mu_0H = $ 0.05 T, and from 0 to 14 T at $T =$ 2 K, on a quantum design physical property measurement system (PPMS DynaCool instrument).

\textbf{Author Contributions}

S.W., Y.H.Z., and H.S.G contributed equally to this work.

\textbf{Notes}

The authors declare no competing financial interest.

\begin{acknowledgement}

Z.C. and Z.L. acknowledge the Key-Area Research and Development Program of Guangdong Province (2020B090922006) and the Key Project of Natural Science Foundation of China (61935010, 61735005).
Z.T. acknowledges the start-up research grant (SRG2016-00002-FST) at the University of Macau and the financial support from the Science and Technology Development Fund, Macao SAR (file no. 063/2016/A2).
D.O. acknowledges the financial support from the Science and Technology Development Fund, Macao SAR (file no. 0029/2018/A1).
H.-F.L. acknowledges the start-up research grant (SRG2016-00091-FST) at the University of Macau, the Guangdong--Hong Kong--Macao Joint Laboratory for Neutron Scattering Science and Technology, and the financial support from the Science and Technology Development Fund, Macao SAR (file nos. 064/2016/A2, 028/2017/A1, and 0051/2019/AFJ).

\end{acknowledgement}

\newpage

\begin{figure*} [!ht]
\centering \includegraphics[width=0.88\textwidth]{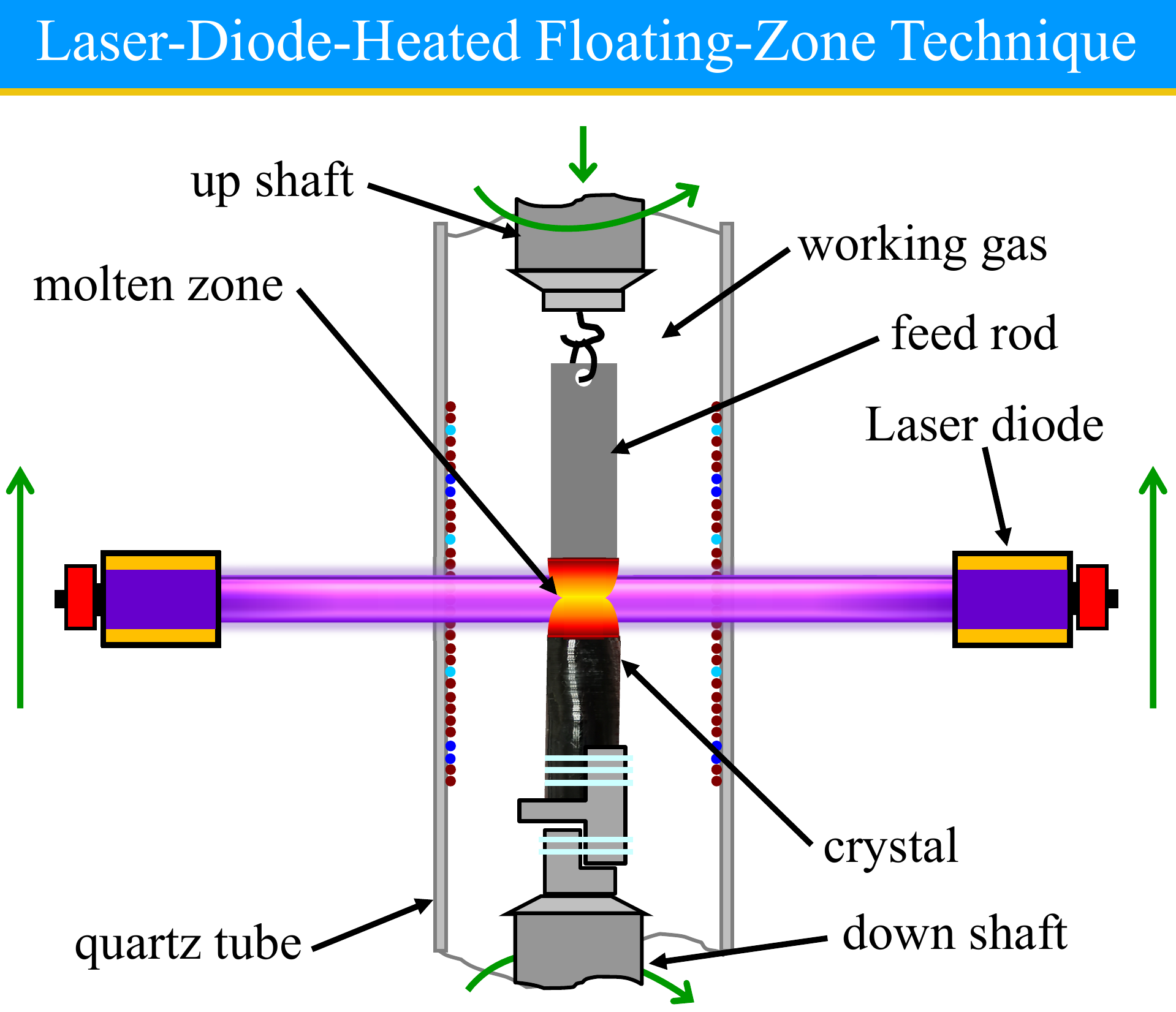}
\caption{For Table of Contents Only}
\label{TOC}
\end{figure*}

\end{document}